\documentclass[preprint,12pt, numbers, compress]{elsarticle}
\pdfoutput=1
\usepackage{amsmath, amsthm, amssymb}
\usepackage{amsfonts}
\usepackage{graphicx}
\usepackage{bm}
\usepackage{textcomp}
\usepackage[normalem]{ulem}

\usepackage{subcaption}
\usepackage{url}
\usepackage{float}
\usepackage{cases}

\usepackage[usenames,dvipsnames]{xcolor}

\usepackage[colorlinks=true, urlcolor=blue, anchorcolor=blue, citecolor=blue,filecolor=blue,linkcolor=blue,menucolor=blue]{hyperref}

\newcommand{\dd}{\mathrm{d}}
\newcommand{\Lap}{\Delta}

\journal{Physica D}

\begin{document}

\begin{frontmatter}

\title{Lattice Brownian bees with cooperative reproduction: steady states, collapse, and spreading}

\author[inst1]{Ohad Vilk}
\ead{ohad.vilk@northwestern.edu}

\author[inst2]{Baruch Meerson}
\ead{meerson@mail.huji.ac.il}

\affiliation[inst1]{
  organization={NSF-Simons National Institute for Theory and Mathematics in Biology},
  city={Chicago},
  state={IL},
  country={USA}
}

\affiliation[inst2]{
  organization={Racah Institute of Physics, Hebrew University of Jerusalem},
  city={Jerusalem},
  postcode={91904},
  country={Israel}
}

\begin{abstract}

We extend the ``Brownian bees'' model of Berestycki et al.\ (2021, 2022) to cooperative reproduction, $kA\to(k{+}1)A$, of a population of $N$ symmetric random walkers with removal, at each birth event, of the particle farthest from the origin. Working in the limit $N\to\infty$, we formulate a hydrodynamic free-boundary problem for this model. Using this formalism, we determine steady state population densities for all~$k$ and prove their linear stability for $k\le 2$ and instability for $k\ge 4$. In the marginal case $k=3$, there is a whole continuous family of steady states at a single critical ratio of the reproduction and diffusion rates. Above criticality the population undergoes an asymptotically self-similar finite-time collapse to the origin. Below the criticality the population spreads diffusively, but the reproduction remains quantitatively relevant. For $k\ge 4$, the unstable steady state separates regimes of a finite-time collapse and diffusive spreading.  Here the collapse dynamics is asymptotically self-similar, and the population density exhibits a scale separation requiring a matched-asymptotic description. Our analytical predictions are confirmed by numerical solutions of the hydrodynamic free-boundary problem and by Monte Carlo simulations of the original microscopic model.

\end{abstract}


\end{frontmatter}

\section{Introduction}
\label{sec:introduction}

$N$-particle branching Brownian motion models with selection form a broad class of stochastic systems in which $N$ particles diffuse, branch, and are removed by a selection rule that keeps the population size fixed~\cite{BD1997,Maillard,DeMasi,BBD,Berestycki1,Berestycki2,VAM2022}. Different rules of particle removal upon birth of a new particle give rise to different macroscopic behaviors, yet they share a common mathematical framework: in the hydrodynamic (HD) limit $N\to\infty$, each of these models reduces to a free-boundary reaction--diffusion problem whose structure encodes the interplay between reproduction and spatial selection. One of the most studied variants is the $N$-BBM model in one dimension, in which the leftmost particle is removed at each branching event~\cite{Maillard,DeMasi,BBD}. This model describes population invasion and predicts a traveling front which belongs to the Fisher--Kolmogorov--Petrovskii--Piscounov (FKPP) universality class~\cite{Fisher1937,KPP,Saarloos2003}. For fronts of this type, microscopic effects --- particle discreteness and finite population size --- lead to a logarithmic in $N$ correction to the mean front speed and large front speed fluctuations ~\cite{BD1997,Derridaetal2006}.

The Brownian bees model~\cite{Berestycki1,Berestycki2} also describes $N$ Brownian particles, each of which can branch into two particles, $A\to 2A$. However, here the population size is kept fixed at~$N$ by removing, at every branching event, the particle \emph{farthest from the origin}. In the HD limit $N\to\infty$ of this model the macroscopic density $u(\mathbf{x},t)$, normalized to unit mass, is governed by a free-boundary problem~\cite{Berestycki1,Berestycki2}:
\begin{align}
&\partial_t u = D\Lap u + \lambda\,u, \quad |\mathbf{x}| < L(t), \label{eq:bb_pde}\\
&u\big|_{|\mathbf{x}|=L(t)} = 0, \quad \int_{|\mathbf{x}|<L(t)} \!\!u(\mathbf{x},t)\,d \mathbf{x}  = 1, \label{eq:bb_bc}
\end{align}
where $D$ is the diffusion coefficient, $\lambda$ is the branching rate, and $\mathbf{x}$ is the position in $d$-dimensional space.  The radius $L(t)$ of the hyper-spherical compact support $|\mathbf{x}|<L(t)$ is determined implicitly by the mass conservation. At long times, this system reaches  a unique spherically symmetric steady state~\cite{Berestycki1,Berestycki2}.
Steady-state fluctuations, including large deviations, of the Brownian bees have been investigated in the framework of fluctuational hydrodynamics and macroscopic fluctuation theory~\cite{MS2021,SSM,SVS}.  Similar models with other selection mechanisms have also been studied~\cite{VAM2022,BarycBees,MV2026}.

In all these works, the population reproduces by simple division, $A\to 2A$. A natural biologically motivated extension is \emph{cooperative} reproduction, $kA\to(k{+}1)A$ with $k\ge 2$, where a successful reproductive event requires the encounter of $k$ individuals at the same location. The binary cooperative reproduction is widespread in nature, but higher-order reproduction also arises in some biological contexts, including quorum-sensing-mediated reproduction in microorganisms~\cite{Miller2001}, multi-gamete fertilization in marine invertebrates~\cite{Levitan1991}, and cooperative breeding in vertebrate and social-insect populations~\cite{CluttonBrock2002}. In the HD description of the reproduction, the $k$-body reproduction rate introduces a nonlinear local density dependence $\sim u^k$. The ensuing nonlinearity for $k>1$ fundamentally changes the spatial structure of the population and, as we shall see, leads to qualitatively new dynamical regimes. The simplest setting for studying such cooperative reproduction is $N$ random walkers on a lattice, where $k$ particles occupying the same site can undergo the branching reproduction $kA\to(k{+}1)A$, and population size is kept fixed by removing one particle at each birth event. Recently, cooperative reproduction of random walkers on a one-dimensional lattice  with a leftmost-particle removal was shown to support, in the HD limit of the model,  robust invasion fronts (mathematically, traveling-wave solutions) only for $k\le 2$ and to lead to nontrivial dynamic behaviors for $k\ge3$~\cite{VM2026}.

Here we generalize the one-dimensional Brownian bees model to cooperative reproduction, $kA\to(k{+}1)A$, of random walkers on a one-dimensional lattice, retaining the farthest-particle removal rule: at each birth event, the particle farthest from the origin (or one such particle if there is more than one) is instantaneously removed, keeping the population size fixed at~$N$. We obtain, for all $k$, exact HD steady-state density profiles. We show that the steady state undergoes a stability transition at $k=3$. For $k\le 2$ the steady state is linearly stable. At $k=3$ the HD problem is scale-invariant, and (a continuous family of) steady states exist only at a single critical value of $\alpha=\lambda/D$. For $k\ge 4$ the steady state is linearly unstable and separates the dynamics between collapse to a point and diffusive spreading. We characterize the marginal $k=3$ dynamics in detail: self-similar collapse for $\alpha>\alpha_c$, self-similar spreading for $\alpha<\alpha_c$, and a continuous family of steady-state density profiles at criticality. For $k=4$ we uncover a scale separation and develop a matched-asymptotic description of the collapse dynamics. Throughout the paper, we compare our predictions with numerical solutions of the HD free-boundary problem and with Monte Carlo (MC) simulations of the underlying microscopic particle system.

The remainder of the paper is organized as follows. In Sec.~\ref{sec:model} we define the model of lattice Brownian bees with cooperative reproduction
and present its HD limit. Section~\ref{sec:results} presents the results: the steady-state framework and exact density profiles (Sec.~\ref{sec:steady}), the linear stability analysis (Sec.~\ref{sec:stability}), and the dynamical regimes for $k=2$ (Sec.~\ref{sec:k2}), $k=3$ (Sec.~\ref{sec:k3}), and $k\ge 4$ (Sec.~\ref{sec:k4}). We conclude with a discussion in Sec.~\ref{sec:discussion}. Implementation details of the HD solver and the MC simulations, the analytical proof of instability for $k>3$, and the derivation of the logarithmic correction to the boundary position for $k=4$ are delegated to the Appendices.

\section{Microscopic model and hydrodynamic theory}
\label{sec:model}
\begin{figure}[t]
\centering
\includegraphics[width=0.9\textwidth]{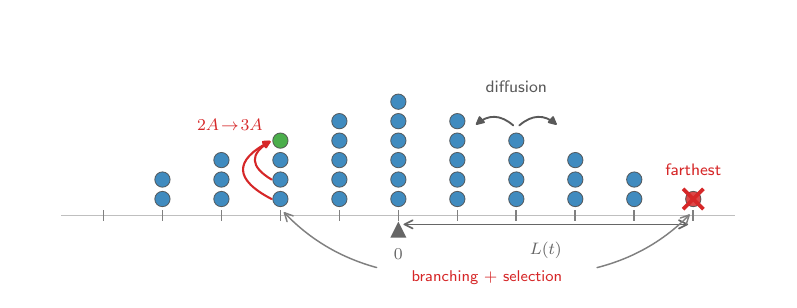}
\caption{Lattice Brownian bees with cooperative reproduction. Particles (circles) on a one-dimensional lattice perform symmetric random walk. When $k$ particles meet at a site, one offspring (green circle) can be produced: $kA\to(k{+}1)A$, shown here for $k=2$. Simultaneously, a particle on the site(s) farthest from the origin (red cross) is removed, keeping the total number of particles~$N$ fixed.}
\label{fig:schematic}
\end{figure}

Our microscopic model involves $N\gg 1$ random walkers on a one-dimensional lattice with lattice constant $h$ and no exclusion constraint, such that multiple occupancy of the same site is allowed. Each particle hops to each of its two neighboring sites with equal rate $\mathcal{D}$. 
At each lattice site, $k$ co-located particles can undergo the cooperative branching reproduction
\begin{equation}\label{eq:rxn}
  kA \;\longrightarrow\; (k{+}1)\,A
\end{equation}
with rate $\Lambda\binom{n}{k}$, where $n$ is the local occupancy and $\Lambda$ is a microscopic rate constant. At each birth event, one of the particles occupying the farthest site from the origin (or any of the two farthest sites if there are two) is instantaneously removed, keeping the total population fixed at~$N$. For $k=1$ this model describes a lattice version of the continuous-in-space Brownian bees model~\cite{Berestycki1,Berestycki2}. The new lattice model is illustrated schematically in Fig.~\ref{fig:schematic}.

In the limit $N\to\infty$, when the local occupancy is large, $n\gg 1$, the binomial rate can be approximated by its leading-order term: $\Lambda\binom{n}{k}\simeq \Lambda\,n^k/k!$~\cite{Gardiner}. In addition, we assume that the ($k$-dependent) characteristic macroscopic length scale of the system is much larger than the lattice spacing $h$, and employ an HD approximation.  Introducing the coarse-grained density $u(x,t)$, normalized so that $\int u\,\dd x = 1$, and taking the continuum limit $h\to 0$ with $D = \mathcal{D}\,h^2$ and $\lambda=\Lambda\,N^{k-1}h^{k-1}/k!$
held fixed,  we arrive at the reaction-diffusion equation
\begin{equation}\label{eq:pde}
  \partial_t u = D\,u'' + \lambda\,u^k\,,
\end{equation}
where primes denote spatial derivatives,  and $D$ is the diffusion coefficient.  The farthest-particle removal imposes absorbing boundary conditions at the edges of a symmetric compact support $|x|<L(t)$:
\begin{equation}\label{eq:bc}
  u\big|_{x= \pm L(t)} = 0\,.
\end{equation}
The support half-width $L(t)$ is determined implicitly by the mass constraint
\begin{equation}\label{eq:mass}
  \int_{-L(t)}^{L(t)} u(x,t)\,\dd x = 1\,.
\end{equation}
This HD formulation, completed by specifying an initial condition $u(x,0)$, is the focus of the present work; we briefly comment on finite-$N$ effects in Sec.~\ref{sec:k2} and in the Discussion. The model generalizes naturally to $d>1$ spatial dimensions, but this generalization is left for future work.

A valuable insight into this HD model is provided by dimensional analysis~\cite{VM2026}. The only dimensional parameters entering Eqs.~\eqref{eq:pde}--\eqref{eq:mass} are $\lambda$, with units $\mathrm{length}^{k-1}/\mathrm{time}$, and $D$, with units $\mathrm{length}^2/\mathrm{time}$. For $k\ne 3$ they define characteristic length and time scales
\begin{equation}\label{eq:scales}
  \ell = \left(\frac{D}{\lambda}\right)^{\!\frac{1}{3-k}}, \qquad
  \tau = \left(\frac{D^{k-1}}{\lambda^2}\right)^{\!\frac{1}{3-k}}.
\end{equation}
Rescaling $x$ by $\ell$, $t$ by $\tau$, and $u$ by $1/\ell$ renders the problem parameter-free; equivalently, we can set $D=\lambda=1$ without loss of generality. Unless physical units are restored explicitly, all results below for $k\ne 3$ are stated in these dimensionless units. For $k=3$ the parameters $\lambda$ and $D$ have the same units, so no intrinsic length or time scale exists, and the dynamics is controlled by the dimensionless ratio $\alpha = \lambda/D$.

\section{Results}
\label{sec:results}

\subsection{Steady states}
\label{sec:steady}

We begin with an analysis of steady-state solutions. Because of their reflection  symmetry $u(-x)=u(x)$, we can restrict our attention to $x\in[0,L]$. The steady-state problem is described by a one-dimensional Lane-Emden equation~(see, e.g., Ref.~\cite{Zwillinger}) with the problem-specific boundary conditions and normalization:
\begin{equation}\label{eq:ss_ode}
  D\,u'' + \lambda\,u^k = 0\,,\qquad
  u'(0) = u(L) = 0\,,\quad
  \int_0^L u\,\dd x = 1/2\,.
\end{equation}
The first integral is
\begin{equation}\label{eq:first_integral}
  (u')^2 = \frac{2\lambda}{D(k+1)}
    \bigl(u_m^{k+1} - u^{k+1}\bigr)\,,
\end{equation}
where $u_m = u(0)$ is the maximum density. Substituting $w = u/u_m$ in $\dd x = -\dd u/|u'|$ and evaluating the resulting integrals for the half-width and the mass gives
\begin{align}
  L &= \sqrt{\frac{D(k+1)}{2\lambda}}\;
    u_m^{(1-k)/2}\;I_1(k)\,,
    \label{eq:L_general} \\
  \frac{1}{2} &=
    \sqrt{\frac{D(k+1)}{2\lambda}}\;
    u_m^{(3-k)/2}\;I_2(k)\,,
    \label{eq:mass_general}
\end{align}
where
\begin{align}
  I_1(k) &= \int_0^1 \frac{\dd w}{\sqrt{1-w^{k+1}}}
    = \frac{1}{k+1}\,B\!\left(\frac{1}{k+1},\,\frac{1}{2}\right),
    \label{eq:I1} \\
  I_2(k) &= \int_0^1 \frac{w\,\dd w}{\sqrt{1-w^{k+1}}}
    = \frac{1}{k+1}\,B\!\left(\frac{2}{k+1},\,\frac{1}{2}\right)
    \label{eq:I2}
\end{align}
and $B(a,b)$ is the Euler beta function~\cite{DLMF}. The density profile itself can be expressed through the regularized incomplete beta function
$I_x(a,b) = B(x;\,a,b)/B(a,b)$:
\begin{equation}\label{eq:profile_general}
  u(x) \! = \! u_m
  \left[I^{-1}_{1-|x|/L}\!\left(\frac{1}{k+1},\,\frac{1}{2}\right)
  \right]^{1/(k+1)},
  \quad x\in[-L,L]\,,
\end{equation}
where $I^{-1}_p(a,b)$ is the inverse of $I_p(a,b)$ with respect to $p$, and $p=1-|x|/L$ in Eq.~\eqref{eq:profile_general}.

The exponent $(3-k)/2$ in the mass constraint~\eqref{eq:mass_general} controls the qualitative behavior of the steady states. For $k<3$ the mass constraint uniquely determines $u_m$, and $L$ follows from~\eqref{eq:L_general}; the steady state is unique. At $k=3$ the exponent vanishes, so the mass constraint fixes the ratio $\lambda/D$ rather than~$u_m$: a steady state exists only at a single critical value
$\alpha_c = \lambda/D$. For the critical value of $\alpha$ the peak density $u_m$ is a free parameter, therefore a whole continuous family of steady states exists. For $k>3$ the mass constraint again determines~$u_m$. This steady state is unique but, as we show below, linearly unstable. 

For $k=1$ (the ordinary Brownian bees), the general expression (\ref{eq:profile_general}) reduces to the known cosine profile $u(x) = \cos(x/\ell)/(2\ell)$ with $\ell = \sqrt{D/\lambda}$ and support radius $L = \pi\ell/2$~\cite{Berestycki1}. The density profiles for $k=2,3$ and $4$ are discussed in detail in the per-$k$ subsections below. Table~\ref{tab:steady} collects the steady-state parameters for $k=1,\ldots,4$.

\begin{table}[t!]
\centering
\begin{tabular}{ccccl}
\hline\hline
$k$ & $\ell$ & $L/\ell$ & $\ell\,u_m$ & Stability \\
\hline
1 & $\sqrt{D/\lambda}$  & $\pi/2$  & $1/2$    & Stable \\
2 & $D/\lambda$          & 3.63     & 0.224    & Stable \\
3 & ---                  & ---      & free     & Marginal$^*$ \\
4 & $\lambda/D$          & 0.157    & 5.41     & Unstable \\
\hline\hline
\end{tabular}
\caption{Parameters of the steady state for $k=1,\ldots,4$. For $k\ne 3$ the rescaled density profile is unique, parameter-free and exists for all $\lambda/D$. $^*$For $k=3$ a continuous family of steady states, parametrized by the peak density $u_m$  (see Sec.~\ref{sec:k3}), exists at  $\alpha\equiv  \lambda/D=\pi^2/2$.}
\label{tab:steady}
\end{table}
\subsection{Linear stability}
\label{sec:stability}
We now analyze the linear stability of the one-dimensional steady states derived in Sec.~\ref{sec:steady}. Consider a small perturbation of the form $u(x,t) = u_s(x) + \varepsilon\,v(x)\,e^{\sigma t}$ with $\varepsilon\ll 1$. Substituting this into Eq.~\eqref{eq:pde} and linearizing, we arrive at a Schr\"odinger-type equation
\begin{equation}\label{eq:eigenvalue}
  D\,v'' + k\lambda\,u_s^{k-1}(x)\,v = \sigma\,v
  \qquad x\in[-L,L]\,,
\end{equation}
with potential $V(x) = -k\lambda\,u_s^{k-1}(x)$ and energy $E=-\sigma$~\cite{CourantHilbert}. With boundary conditions which we will discuss shortly, this equation defines an eigenvalue problem. The steady state is unstable if any eigenvalue $\sigma$ is positive.

To determine the boundary conditions for $v(x)$, we note that the support of $u(x,t)$ must remain symmetric, so the boundary shifts as $L\to L + \varepsilon\,\delta L\,e^{\sigma t}$ on both sides simultaneously. Taylor-expanding the absorbing boundary condition at both endpoints $x = \pm L$ and requiring consistency yields the condition
\begin{equation}\label{eq:bc_compat}
  v(L) = v(-L)\,.
\end{equation}
The boundary displacement is then determined by $\delta L = -v(L)/u_s'(L)$, and mass conservation at first order requires
\begin{equation}\label{eq:mass_pert}
  \int_{-L}^{L} v\,\dd x = 0\,.
\end{equation}
By the symmetry $V(-x)=V(x)$ of the potential, the eigenmodes decompose into even modes $v(-x)=v(x)$ and odd modes $v(-x)=-v(x)$. For even modes, the endpoint condition~\eqref{eq:bc_compat} is automatically satisfied, the boundary displacement $\delta L$ is free, and the mass constraint~\eqref{eq:mass_pert} provides an additional condition. For odd modes, condition~\eqref{eq:bc_compat} forces $v(L)=0$, while the mass constraint is automatically satisfied by antisymmetry; the admissible problem on $0<x<L$ therefore has Dirichlet boundary conditions $v(0)=v(L)=0$. 
In this latter case, differentiating the steady-state equation with respect to $x$, we see that $u_s'$ solves Eq.~\eqref{eq:eigenvalue} with $\sigma=0$, as it satisfies the left condition $u_s'(0)=0$; however, it has no zeros in $(0,L]$. Thus, at $\sigma=0$, the first positive zero of the odd unconstrained solution is outside the interval $(0,L]$. By Sturm's comparison theorem \cite{coddington1956theory}, to make this zero occur at $x=L$ one must decrease $\sigma$; hence the largest odd eigenvalue is strictly negative, ruling out the odd sector as a source of instability. 

In the even sector, for $k=1$ (the ordinary Brownian bees) the eigenvalues constrained by Eq.~\eqref{eq:mass_pert} can be easily computed analytically: $\sigma_n = \lambda(1-4n^2)$ for $n=1,2,\ldots$, with eigenfunctions $\sim\cos(2n\,x/\ell)$. Since all the eigenvalues are negative, the steady state is linearly stable.
For general~$k$, the leading even eigenvalue~$\sigma_0$, constrained by Eq.~\eqref{eq:mass_pert}, can be found numerically by a shooting method~\cite{shooting}: for a trial value of~$\sigma$, one integrates Eq.~\eqref{eq:eigenvalue} from $x=0$ (with $v'(0)=0$, $v(0)=1$) to $x=L$ and adjusts~$\sigma$ until the mass constraint~\eqref{eq:mass_pert} is satisfied. The result is that for $k< 3$ the steady state is stable ($\sigma_0<0$),  while for $k> 3$ it is unstable ($\sigma_0>0$). Notably, in the latter case the instability can also be  established analytically via a mass-slope criterion, as we show in~\ref{app:instability}.

\subsection{\texorpdfstring{$k=2$}{k=2}: stable steady state}
\label{sec:k2}

The characteristic macroscopic length for $k=2$ is $\ell = D/\lambda$. Rescaling $\xi = x/\ell$, $\phi = \ell\,u$ brings the steady-state equation into the parameter-free form $\phi'' + \phi^2 = 0$ with $\phi'(0)=0$ and $\phi(\rho)=0$, where $\rho=L/\ell$. The first integral gives
$(\phi')^2 = (2/3)(\phi_m^3-\phi^3)$, and Eqs.~\eqref{eq:L_general} and \eqref{eq:mass_general} yield $\phi_m\simeq 0.224$ and $\rho \simeq 3.63$. In physical units, $u_m = \phi_m\,\lambda/D$ and $L = \rho\,D/\lambda$. The density profile can also be expressed through the Weierstrass elliptic function~\cite{WhittakerWatson} (which takes a simplified form since the first integral is purely cubic):
\begin{equation}\label{eq:profile_k2}
  \phi(\xi) = -6\,\wp\!\bigl(\xi + \omega;\;0,\;-\phi_m^3/54\bigr)\,,
\end{equation}
where $\omega$ is the half-period at which $\wp'(\omega)=0$.

\begin{figure}[t]
\centering
\includegraphics[width=\textwidth]{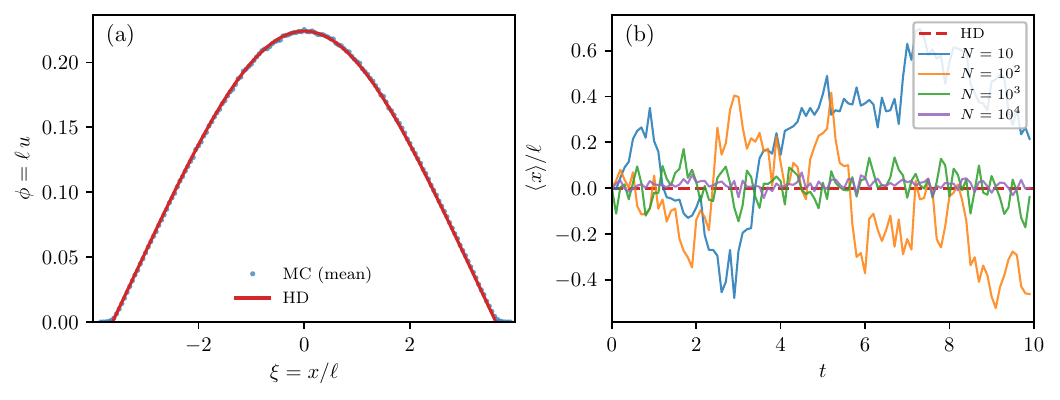}
\caption{Rescaled steady-state observables for $k=2$ with $\ell=20$. (a)~Density profile $\phi=\ell u$ versus $\xi=x/\ell$: $N=10^4$, $D=10^5$, $\Lambda=h=1$; circles = MC ensemble mean (100 realizations); solid = HD~\eqref{eq:profile_general}. (b)~Center-of-mass position $\langle x\rangle/\ell$ versus time; dashed = HD ($\langle x\rangle=0$); colored curves = single stochastic realizations at $N=10$, $10^2$, $10^3$, and $10^4$.}
\label{fig:k2_profile}
\end{figure}

Figure~\ref{fig:k2_profile}(a) compares the analytical density profile with our MC simulations. With $N=10^4$ particles, $D=10^5$, and $\Lambda=h=1$ (giving $\lambda=5 \times 10^3$ and $\ell=20$), the rescaled density, measured in the simulations and averaged over 100 realizations, is in excellent agreement with the theoretical curve.  This agreement supports the validity of the HD limit and confirms the stability of the $k=2$ steady state. Figure~\ref{fig:k2_profile}(b) shows the rescaled center-of-mass position $\langle x\rangle/\ell$ as a function of time: the HD prediction $\langle x\rangle=0$ (dashed) and individual stochastic realizations at $N=10$, $10^2$, $10^3$, and $10^4$.

\subsection{\texorpdfstring{$k=3$}{k=3}: marginal stability}
\label{sec:k3}
With $I_2(3) = \pi/4$, the mass constraint~\eqref{eq:mass_general} becomes
\begin{equation}\label{eq:k3_alpha_c}
  \frac{1}{2} = \sqrt{\frac{2D}{\lambda}}\;\frac{\pi}{4}
  \qquad\Longrightarrow\qquad
  \alpha_c \equiv \frac{\lambda}{D} = \frac{\pi^2}{2}
    \simeq 4.935\,.
\end{equation}
At this single critical value of~$\alpha$, a whole continuous family of steady states exists, parametrized by the peak density $u_m$. The half-width adjusts so that
$L u_m = K(1/2)/\sqrt{\alpha_c}$, where $K(m)$ is the complete elliptic integral of the first kind
($K(1/2)\simeq 1.854$). The density profile has the closed form
\begin{equation}\label{eq:profile_k3}
  u(x) = u_m\,\operatorname{cn}\!\left(
    u_m\sqrt{\alpha_c}\;x \;\Big|\; 1/2\right),
\end{equation}
where $\operatorname{cn}(z\mid m)$ is the Jacobi elliptic cosine~\cite{DLMF,WhittakerWatson}. This follows from the identity
$\operatorname{cn}''(z\mid m) = (2m{-}1)\operatorname{cn} - 2m\operatorname{cn}^3$,
which at $m=1/2$ reduces to $\operatorname{cn}''(\dots) = -\operatorname{cn}^3(\dots)$. In Fig.~\ref{fig:k3_regimes}(c,d) we compare Eq. \eqref{eq:profile_k3} to HD numerics and MC simulations.

For $\alpha\ne\alpha_c$ no steady state exists that complies with the mass constraint. The population dynamics for both $\alpha < \alpha_c$ and $\alpha > \alpha_c$ is asymptotically described by self-similar solutions, in which diffusion and reproduction remain of the same order of magnitude --- a direct consequence of $k=3$ being the mass-critical exponent. 
By virtue of the mass-preserving rescaling $u \sim L^{-1}$, $x\sim L$, the self-similar form of $u(x,t)$ is
\begin{equation}\label{eq:ss_ansatz}
  u(x,t) = \frac{1}{L(t)}\,F\!\left(\frac{x}{L(t)}\right),
  \qquad 0 < y = \frac{x}{L(t)} < 1\,,
\end{equation}
where the scale function $F$ satisfies
\begin{equation}\label{eq:ss_bcs}
  F'(0) = 0\,,\quad F(1) = 0\,,\quad
  \int_0^1 F(y)\,\dd y = 1/2\,,
\end{equation}
and we have used the reflection symmetry of the scale function $F$. In Sec.~\ref{sec:k3_collapse} we study the finite-time collapse for $\alpha > \alpha_c$, and in Sec.~\ref{sec:k3_expansion} the self-similar expansion for $\alpha < \alpha_c$.

\subsubsection{Self-similar collapse (\texorpdfstring{$\alpha>\alpha_c$}{alpha>alpha\_c})}
\label{sec:k3_collapse}

For $\alpha > \alpha_c$ the support shrinks to zero in finite time: $L(t)\to 0$ as $t\to T$. Substituting the ansatz~\eqref{eq:ss_ansatz} into~\eqref{eq:pde} and defining the dimensionless collapse rate $\gamma = -\dot{L}L/D > 0$ we arrive at the nonlinear eigenvalue problem
\begin{equation}\label{eq:k3_collapse_ode}
  F'' + \alpha\,F^3 = \gamma\bigl(F + y\,F'\bigr)\,,
  \qquad 0 < y < 1\,,
\end{equation}
with the boundary and mass conditions~\eqref{eq:ss_bcs}. For each $\alpha > \alpha_c$, the function $F$ and the eigenvalue $\gamma$ can be determined simultaneously by a shooting method. Assuming that $\gamma$ is constant near the blowup time, and integrating the equation $\dot{L}\,L = -D\gamma$, we obtain
\begin{equation}\label{eq:k3_collapse_scaling}
  L(t) \sim \sqrt{2D\gamma\,(T-t)}\,,
  \qquad
  u(0,t) \sim \frac{F(0)}{\sqrt{2D\gamma}}\,(T-t)^{-1/2}\,.
\end{equation}
The dynamical scalings of the collapse are therefore $L\sim\tau^{1/2}$ and $u_{\max}\sim\tau^{-1/2}$, where $\tau = T-t$.

\begin{figure}[t]
\centering
\includegraphics[width=\textwidth]{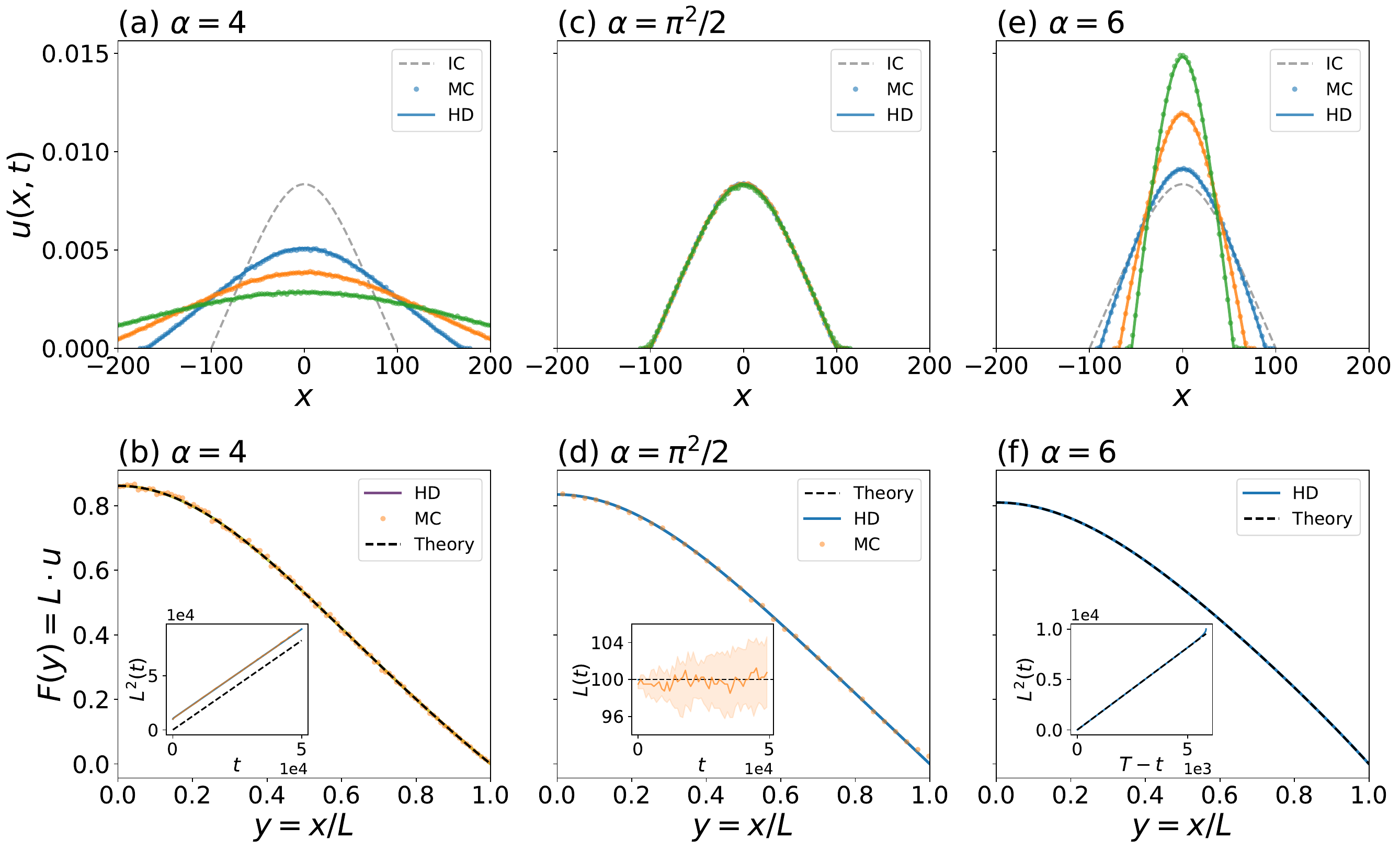}
\caption{Population dynamics for $k=3$ in the three regimes: expansion ($\alpha = 4$, panels a,b), steady state ($\alpha = \alpha_c = \pi^2/2$, panels c,d), and collapse ($\alpha = 6$, panels e,f).
(a,c,e)~Density profiles $u(x,t)$ in physical coordinates as functions of $x$ at successive times. Dashed gray line (clearly visible on panels a and c) shows the initial condition: the Jacobi cn-steady state \eqref{eq:profile_k3} at half-width $L_0 = 100$. Colored solid lines: numerical HD solutions at $t = 11{,}000,\; 24{,}000,\; 49{,}000$ (a,c) and $t = 1{,}000,\; 3{,}000,\; 4{,}000$ (e). Colored circles: MC simulations ($N = 5\times 10^4$ particles, averaged over 20 independent runs) at matching times. 
(b,d,f)~Density scaling function $F(y) = L u$ vs.  $y = x/L$. Dashed curves: numerical solutions of Eqs.~\eqref{eq:k3_expansion_ode} (b) and~\eqref{eq:k3_collapse_ode} (f). Solid curves: HD numerics. Circles: MC simulations.
Insets in (b) and (f): $L^2(t)$ versus time; dashed lines show the predicted $2\beta t$ scaling (b, $\beta\simeq 0.82$) and $2\gamma(T-t)$ scaling (f, $\gamma\simeq 0.82$, $T \simeq 5{,}830$). Inset in (d): $L(t)$ fluctuates around $L = L_0 = 100$. We set $D = h = 1$.}
\label{fig:k3_regimes}
\end{figure}

Figure~\ref{fig:k3_regimes}(e,f) compares the scaling function $F$, obtained by numerically solving  Eq.~\eqref{eq:k3_collapse_ode} with a numerical solution of the complete time-dependent free-boundary problem for $\alpha = 6$ ($\gamma \simeq 0.82$). The blowup time $T$ depends on the initial condition and is determined from the HD numerics by fitting the late-time scaling $L^2(t) \simeq 2D\gamma(T - t)$. The inset in panel~(f) confirms the predicted scaling $L^2(t) \propto T-t$.

\subsubsection{Self-similar expansion (\texorpdfstring{$\alpha<\alpha_c$}{alpha<alpha\_c})}
\label{sec:k3_expansion}

For $\alpha < \alpha_c$ the reproduction is too weak to sustain a finite-width steady state, and the support expands: $L(t)\to\infty$ as $t\to\infty$. The same mass-preserving ansatz~\eqref{eq:ss_ansatz} applies. Defining $\beta = \dot{L}\,L/D > 0$ for the spreading case, substitution into~\eqref{eq:pde} gives
\begin{equation}\label{eq:k3_expansion_ode}
  F'' + \alpha\,F^3 + \beta\bigl(F + y\,F'\bigr) = 0\,,
  \qquad 0 < y < 1\,,
\end{equation}
again with conditions~\eqref{eq:ss_bcs}.  For each $\alpha<\alpha_c$, the eigenvalue $\beta$ and the scaling function $F$ can be determined by a shooting method: one integrates~\eqref{eq:k3_expansion_ode} from $y=0$ with $F'(0)=0$ and a trial value of $\beta$, treating $F(0)$ as a free parameter set by the normalization $\int_0^1 F\,\dd y = 1/2$; the shooting parameter $\beta$ is then adjusted until the boundary condition $F(1)=0$ is satisfied. Integrating $\dot{L}\,L = D\beta$ at long times gives
\begin{equation}\label{eq:k3_expansion_scaling}
  L(t) \sim \sqrt{2D\beta\,t}\,,
  \qquad
  u(0,t) \sim \frac{F(0)}{\sqrt{2D\beta}}\,t^{-1/2}\,.
\end{equation}
The long-time spreading is diffusive, $L\sim t^{1/2}$, but it is not free diffusion: the cubic  reproduction term remains quantitatively relevant, so the scaling function $F(y)$ in Eq. (\ref{eq:k3_expansion_ode})  is  non-Gaussian and depends on~$\alpha$.

Figure~\ref{fig:k3_regimes}(a,b) shows the self-similar expansion for $\alpha = 4$ ($\beta \simeq 0.82$), compared with the HD numerics and MC simulations. The inset in panel~(b) confirms  the long-time behavior $L^2(t) \propto t$.

\begin{figure}[t!]
\centering
\includegraphics[width=\textwidth]{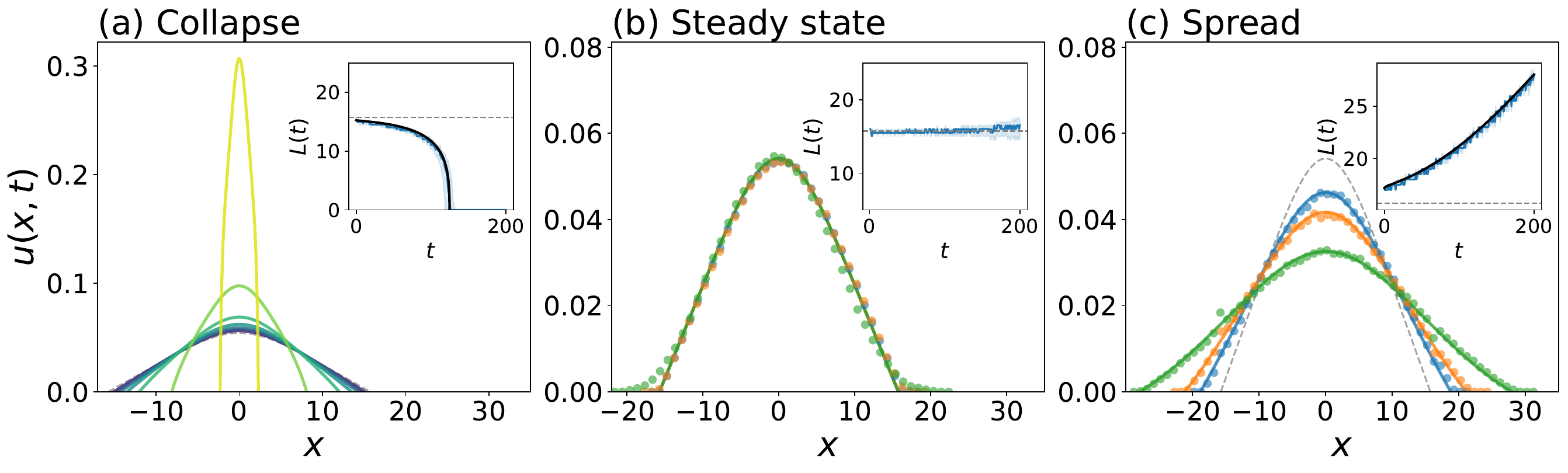}
\caption{Population dynamics for $k=4$, starting from rescaled copies of unstable steady-state profile~\eqref{eq:profile_general} with different initial half-widths $L(0)$. Dashed gray curve (clearly visible only on panel c): steady-state profile. Insets show $L(t)$: blue band = MC interquartile range (25th--75th percentile), black curve = HD, dashed gray = steady-state value $L_{\mathrm{SS}}\simeq 15.7$.
(a)~Collapse (narrow IC, $L(0) = 15.3$): HD density profiles at seven times from $10\%$ to $99.9\%$ of the numerically found blowup time $T \simeq 125$ (dark to light).
(b)~Steady state ($L(0) = L_{\mathrm{SS}}$): solid curves = HD, markers = MC (20 realizations, averaged); $t = 5,\,10,\,20$.
(c)~Spread (wide IC, $L(0) = 17.2$): same convention as~(b); $t = 5,\,10,\,20$. The parameters are $N=5\times 10^4$, $D=h=1$, $\lambda=100$, $\ell=\lambda/D=100$.}
\label{fig:k4_profiles}
\end{figure}

\subsection{\texorpdfstring{$k\ge 4$}{k>=4}: unstable steady state, collapse and spreading}
\label{sec:k4}
For $k\ge 4$ the steady state is linearly unstable (Sec.~\ref{sec:stability}). Qualitatively, it acts as a separatrix in a function space: initial conditions narrower than the steady state collapse to a strongly localized configuration (within the HD description to a point), while broader initial conditions spread diffusively as the $u^k$ reproduction term becomes subleading. Figure~\ref{fig:k4_profiles} demonstrates this behavior for $k=4$ using HD numerics and MC simulations: starting from rescaled copies of the analytical density profile with different initial half-widths $L(0)$, the population either collapses or spreads, depending (roughly speaking) on which side of the steady state the initial condition lies.

Now we consider the case of $k=4$ in some detail, and we start from the unstable steady state.  The characteristic length scale here is $\ell = \lambda/D$, and the rescaled steady-state equation $\phi'' + \phi^4 = 0$ is parameter-free, with $\phi_m \simeq 5.41$ and $\rho \simeq 0.157$. In physical units,
$u_m = \phi_m\,D/\lambda$ and $L = \rho\,\lambda/D$. The resulting (unstable) steady-state density profile is very peaked and narrow.

\subsubsection{$k=4$: Matched-asymptotic description of collapse}
\label{sec:k4_collapse}

The key observation is that, as the collapse develops, the diffusion term becomes subleading over most of the support $|x|<L(t)$; this property is in contrast to $k=3$, where diffusion remains relevant. Under the mass-preserving scaling $u\sim L^{-1}$ and $x\sim L$, the two terms on the right-hand side of Eq.~\eqref{eq:pde} behave as $\lambda u^4\sim L^{-4}$ and $D u_{xx}\sim L^{-3}$. As $L\to 0$, reproduction dominates in the bulk and diffusion becomes subleading; the time derivative $u_t$ then follows the same $L^{-4}$ balance.

This argument, however, does not apply near the edge of support $|x|=L$, where $u$ must vanish. As a result, the reproduction dominates in the bulk (the outer region), while the diffusion dominates in narrow boundary layers near $|x|=L$, whose size goes to zero faster than $L(t)$.  The ensuing length scale separation 
necessitates a matched-asymptotic expansion~\cite{BenderOrszag}, which we will now present, again exploiting the mirror-symmetry of $u(x,t)$ around the origin.

Let us denote for convenience $\varepsilon = L(t) \to 0$ and introduce the outer coordinate $y=x/\varepsilon \in [0,1]$. The density $u(x,t)$ in the outer region can be written as
\begin{equation}\label{eq:k4_outer_def}
  u(x,t) = \frac{1}{\varepsilon}\,A(y,\varepsilon)\,,
\end{equation}
where the factor $1/\varepsilon$ is enforced by the mass constraint $\int_0^1 A\,\dd y = 1/2$. Substituting this into Eq.~\eqref{eq:pde} and multiplying by $\varepsilon^4$, we obtain the exact outer-region equation
\begin{equation}\label{eq:k4_outer_exact}
  \mu\bigl(A + y\,A_y\bigr) - \mu\varepsilon\,A_\varepsilon
  = \lambda\,A^4 + \varepsilon\,D A_{yy}\,,
\end{equation}
where we have defined the collapse rate $\mu= -\,\varepsilon^2\dot{\varepsilon}$. The diffusion term in Eq.~(\ref{eq:k4_outer_exact}) appears with a prefactor $\varepsilon$, confirming that it is subleading in the outer region.

Expanding $A(y, \varepsilon)= F_0(y) + \varepsilon\,F_1(y) + O(\varepsilon^2)$ and $\mu = \mu_0 + \varepsilon\,\mu_1 + O(\varepsilon^2)$, we obtain the leading-order outer-region equation
\begin{equation}\label{eq:k4_outer_leading}
  \mu_0\bigl(y\,F_0\bigr)' = \lambda\,F_0^4\,.
\end{equation}

Solving this equation and demanding the positivity and the mass condition [which in the leading order is simply $\int_0^1 F_0\,\dd y = 1/2$], we arrive at a continuous one-parameter family of outer solutions. 
Introducing a shape parameter~$r$, this family of solutions can be written as
\begin{equation}\label{eq:k4_F0}
  F_0(y) = \frac{1}{2\,I(r)}\bigl(1+r\,y^3\bigr)^{-1/3}\,,
\end{equation}
where
\begin{equation}\label{eq:k4_I_r}
  I(r) = \int_0^1 \bigl(1+r\,y^3\bigr)^{-1/3}\,\dd y = \,
   _2F_1\left(\frac{1}{3}, \frac{1}{3};\frac{4}{3};-r\right)\,,
\end{equation}
and $_2F_1(...)$ is the hypergeometric function. The symmetry condition $\partial_x u|_{x=0} = 0$ is satisfied automatically. 

The leading-order collapse rate as a function of $r$ is
\begin{equation}\label{eq:k4_mu0}
  \mu_0(r) = \lambda\bigl[2\,I(r)\bigr]^{-3}\,.
\end{equation}
Integrating the relation $\mu_0 = -\dot{L}\,L^2$, we obtain the leading-order time-dependence of the support size,
\begin{equation}\label{eq:k4_L_leading}
  L(t) \sim \bigl[3\mu_0(T-t)\bigr]^{1/3}\,,
\end{equation}
and the peak-density 
\begin{equation}\label{eq:k4_peak}
  u(0,t)^{-3} \sim 3\lambda(T-t)\,,
\end{equation}
near collapse. Note that the latter result does not depend on the parameter ~$r$. The predicted dynamical scaling  near collapse, $L\sim\tau^{1/3}$ and $u_{\max}\sim\tau^{-1/3}$, where $\tau = T-t$, is markedly different from the $\tau^{1/2}$ scaling of the mass-critical $k=3$ case.

Notably, the leading-order outer solution does not vanish at $y=1$: $F_b := F_0(1) = (1+r)^{-1/3}/[2\,I(r)] > 0$. The absorbing boundary condition is enforced in a narrow diffusive boundary layer near $x=L(t)$ which we will now consider. Introduce the boundary-layer coordinate
\begin{equation}\label{eq:k4_z}
  z = \frac{L(t) - x}{L(t)^2}\,,
\end{equation}
so the physical layer thickness is $\delta_x \sim L(t)^2$.
Writing $u = (1/\varepsilon)\,B(z,\varepsilon)$ with $B = H_0(z) + \varepsilon\,H_1(z) + \cdots$, the leading boundary-layer equation is
\begin{equation}\label{eq:k4_H0_ode}
  D\,H_0'' + \mu_0\,H_0' = 0\,,
\end{equation}
with $H_0(0) = 0$ (absorbing boundary) and $H_0(\infty) = F_b$ (leading-order matching to the outer solution). 
As one can see,  the boundary layer is purely diffusive in the leading order: the reproduction term is absent, and the dominant balance is between the moving-boundary advection and diffusion. The solution of Eq.~(\ref{eq:k4_H0_ode}) is
\begin{equation}\label{eq:k4_H0}
  H_0(z) = F_b\bigl(1 - e^{-\mu_0 z/D}\bigr)\,,
\end{equation}
which completes the leading-order description of the problem.

\begin{figure}[t!]
\centering
\includegraphics[width=0.7\textwidth]{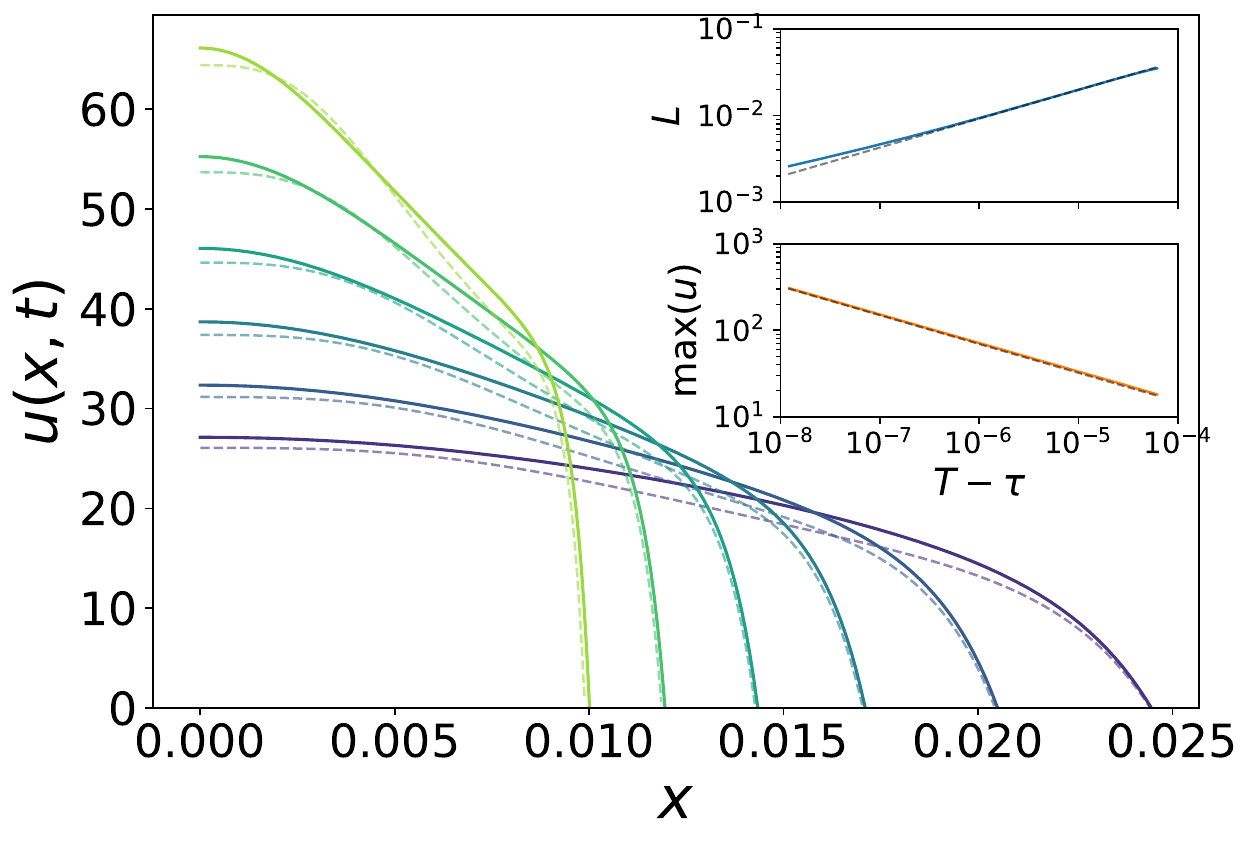}
\caption{Self-similar collapse for $k=4$ near the blowup time. Solid curves: HD density profiles $u(x)$ at six logarithmically spaced times approaching the blowup time~$T$ (from dark to light); dashed curves: matched-asymptotic theory predictions~\eqref{eq:k4_composite_outer} and \eqref{eq:k4_composite_BL} at the same times. Insets: the half-width of support $L$ (top) and $\max(u)$ (bottom) versus $\tau = T-t$ on log-log scale; solid = HD, dashed = leading-order scaling $L\sim(3\mu_0\tau)^{1/3}$~\eqref{eq:k4_L_leading} and $\max(u)\sim(3\lambda\tau)^{-1/3}$~\eqref{eq:k4_peak}. The initial condition is the steady-state profile~\eqref{eq:profile_general} rescaled to half-width $L(0) = 0.035$; the numerically determined blowup time is $T \simeq 6.2\times 10^{-5}$.}
\label{fig:k4_collapse}
\end{figure}

In summary, the density of collapsing population has the two-region composite structure
\begin{align}
  u(x,t) &\sim \frac{1}{L}\,F_0\!\left(\frac{x}{L}\right),
  &&0 \le x < L - O(L^2)\,,\label{eq:k4_composite_outer}\\
  u(x,t) &\sim \frac{1}{L}\,H_0\!\left(\frac{L-x}{L^2}\right),
  &&L - x = O(L^2)\,,\label{eq:k4_composite_BL}
\end{align}
where $L(t) \sim [3\mu_0(T-t)]^{1/3}$. 
The two asymptotics match in their joint validity region $L^2(t)\ll L(t)-x \ll L(t)$.   In the subleading order, which we will not deal with here, the diffusion introduces a subleading correction to the outer solution, while the reproduction introduces a subleading correction to the boundary-layer solution. 

Importantly, the matched-asymptotic solution does not provide a selection rule
for the parameter $r>0$, present in the solution. This parameter is determined by the initial condition. The blowup time~$T$ also depends on the initial condition and is determined numerically by fitting the late-time scaling $[\max u]^{-3} \simeq 3\lambda(T-t)$~\eqref{eq:k4_peak}. Figure~\ref{fig:k4_collapse} compares the asymptotic structure (\ref{eq:k4_composite_outer}) and (\ref{eq:k4_composite_BL}) with HD numerics, showing a good agreement.

\subsubsection{Diffusive spreading}
\label{sec:k4_spreading}

When the initial condition is effectively broader than the unstable steady state, the 
$\lambda u^4$ reproduction term becomes negligible in the bulk of the population as the population spreads. The long-time dynamics of the bulk population is then governed by free diffusion:
\begin{equation}\label{eq:k4_spreading}
  u(x,t) \to \frac{1}{\sqrt{4\pi D\,t}}\,\exp\!\left(-\frac{x^2}{4D\,t}\right)\,.
\end{equation}
Importantly, $L(t)$ grows faster than diffusively:
\begin{equation}\label{eq:k4_Lspreading}
      L(t) \sim \sqrt{2D\,t\,\ln t}\,, \quad \ln t \gg 1.
\end{equation}
The logarithmic correction is determined by the balance between the integrated reproduction term and the absorbing flux to the boundaries (see~\ref{app:log_correction} for a derivation), and both these terms become very small at long times. Note that this spreading regime is different from its counterpart in the mass-critical $k=3$ case, where the cubic reproduction term remains part of the leading-order dynamics.

\begin{figure}[t!]
\centering
\includegraphics[width=\textwidth]{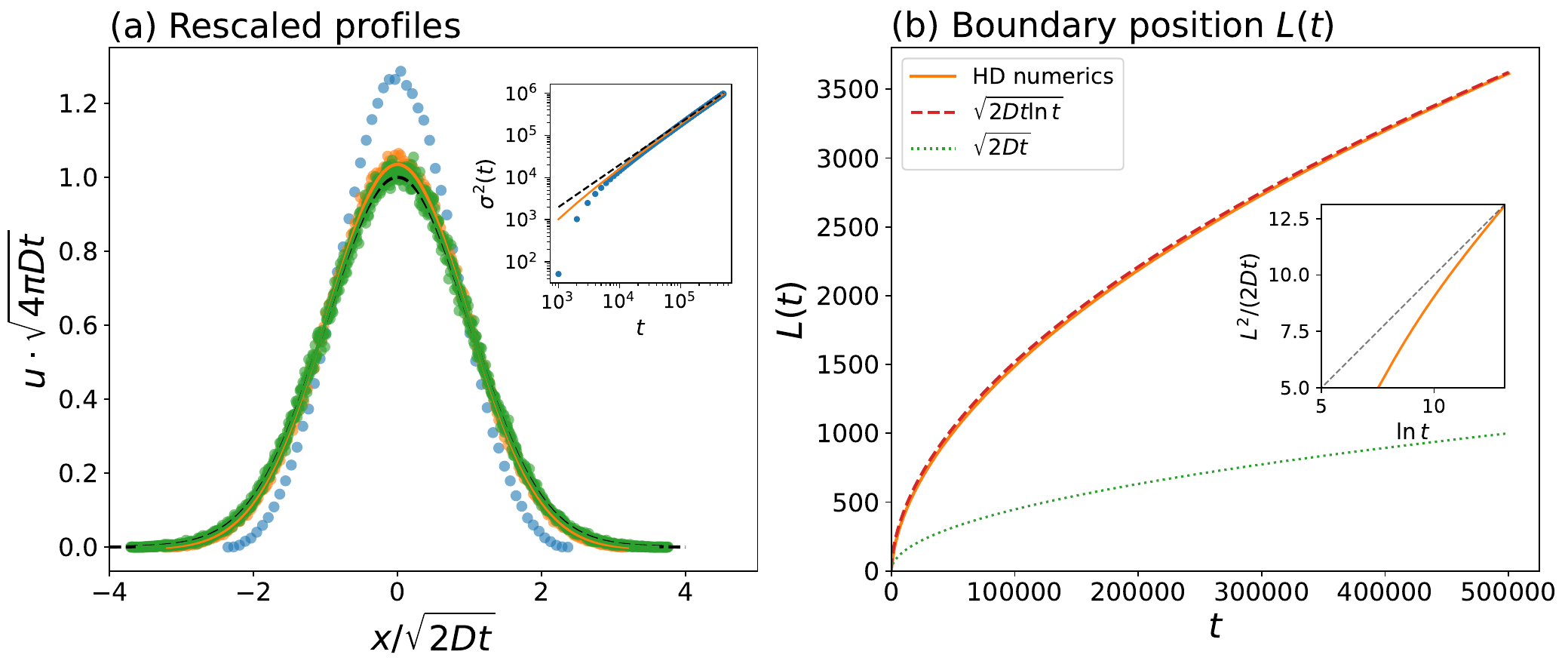}
\caption{Diffusive spreading for $k=4$. Shown are
(a)~Rescaled density profiles, $u\sqrt{4\pi D t}$ vs. $x/\sqrt{2Dt}$: markers = MC ($t=5{\times}10^3,\; 5{\times}10^4,\; 5{\times}10^5$; 10 realizations), solid lines = HD numerics ($t=5{\times}10^4$ and $5{\times}10^5$); dashed line: the Gaussian ~\eqref{eq:k4_spreading}. Inset: variance $\sigma^2(t)$ (circles = MC, solid = HD) vs. the purely diffusive scaling $2Dt$ (dashed).
(b)~Boundary position $L(t)$ from HD numerics (solid) compared with theoretical prediction $\sqrt{2Dt\ln t}$ (dashed  line) and with the pure-diffusion scaling $\sqrt{2Dt}$ (dotted line). Inset: $L^2(t)/(2Dt)$ versus $\ln t$, verifying the logarithmic correction to diffusive scaling of $L(t)$.}
\label{fig:k4_spreading}
\end{figure}

MC simulations confirm Eqs.~(\ref{eq:k4_spreading}) and (\ref{eq:k4_Lspreading}), see Figs.~\ref{fig:k4_profiles}(c) and \ref{fig:k4_spreading}: the bulk variance grows purely diffusively as $\sigma^2 \sim 2Dt$ as the population density asymptotically approaches a Gaussian [Fig. \ref{fig:k4_spreading}(a)], while the boundary position $L(t)$ grows as $\sqrt{2Dt\ln t}$ [Fig. \ref{fig:k4_spreading}(b)]. Figure~\ref{fig:k4_spreading}(b) presents a detailed comparison of $L(t)$ from HD numerics with the asymptotic prediction $\sqrt{2Dt\ln t}$; the inset confirms the logarithmic correction at $\ln t\gg 1$ by plotting $L^2/(2Dt)$ versus $\ln t$.

\section{Discussion}
\label{sec:discussion}

We have extended the Brownian bees model to cooperative reproduction, $kA\to(k{+}1)A$, of $N\gg 1$ symmetric random walkers. We formulated a hydrodynamic (HD) model in the limit of $N\to \infty$ and studied its steady-state and dynamical properties. One central result is a linear stability transition at $k=3$: the steady state is stable for $k<3$, and unstable for $k>3$, see Sec.~\ref{sec:stability}. For $k>3$, instability is proved analytically via a mass-slope criterion, see~\ref{app:instability}. 

In the special case $k=3$ there are no steady states except at a single critical value $\alpha_c = \pi^2/2$ of the dimensionless ratio $\alpha = \lambda/D$, where a whole continuous family of steady states exists. Below the critical value the population spreads, showing an asymptotically self-similar diffusive scaling with time, although the reproduction remains important for all times. Above the criticality the population, as described by HD theory, exhibits an asymptotically self-similar finite-time collapse into a point. The collapse rate and the expansion rate are determined by a nonlinear eigenvalue problem.

For $k\ge 4$ the unstable steady state separates two qualitatively different dynamical regimes: a finite-time self-similar collapse to a point, and a diffusive spreading which is asymptotically Gaussian in the bulk. The collapse dynamics -- which we considered in some detail for $k=4$ --
exhibits a length-scale separation: a reproduction-dominated bulk region of the collapsing population coexists with a narrow diffusion-dominated boundary layer near the edges of the shrinking compact support.  Ultimately, the collapse is arrested by lattice effects, unaccounted for by the continuous HD theory. While these results are illustrated above only for $k=4$,  we believe they hold for any $k > 3$. For $k=5$ we checked this explicitly in MC simulations.

Our analytical predictions are confirmed quantitatively by numerical solutions of the HD free-boundary problem and by Monte Carlo simulations of the underlying microscopic model.

The present work continues the direction set in Ref.~\cite{VM2026}, where cooperative reproduction of random walkers with the \emph{leftmost}-particle removal upon birth was studied. For $k\le 2$, the  leftmost-particle removal rule facilitates invasion fronts -- traveling-wave solutions, which have no counterpart in the present model. Despite this fundamental difference, the two selection rules share a common instability threshold at $k=3$, which reflects the same dimensional-analysis origin: $k=3$ is the mass-critical exponent in one spatial dimension. It would be interesting to determine whether this universality extends to other selection rules, such as the random-particle removal protocol recently studied for $k=1$~\cite{MS2021}, or to nearest-neighbor competition models.

Several natural extensions of this work present themselves. First, it would be interesting to extend the analysis to $d>1$ spatial dimensions. 
Second, the HD description neglects fluctuations due to finite~$N$. For example, the center-of-mass position $\langle x\rangle$ fluctuates around zero, with larger fluctuations at smaller $N$, as illustrated in Fig.~\ref{fig:k2_profile}(b).
For $k=1$, typical fluctuations of the center of mass and of the rightmost particle position were studied in Ref.~\cite{SSM} in the framework of fluctuating hydrodynamics.  It would be very interesting to use the fluctuating hydrodynamics framework for the lattice Brownian bees with $k\ge 2$, especially in the critical regime $k=3$,  where the HD steady states are marginally stable. 

From a broader perspective, the results of this work reinforce a theme already apparent in the leftmost-particle removal model~\cite{VM2026}: somewhat surprisingly, minimal models of population dynamics with spatial selection already encode a strong preference for low-order reproduction. For $k\le 2$ with farthest-particle removal, the population self-organizes into a stable macroscopic collective -- a robust steady state that recovers from perturbations and persists indefinitely. For $k\ge 4$, no such stable configuration exists: the population either collapses to a point (resolved by finite-size effects) or spreads diffusively. The marginal case $k=3$ sits precisely at the boundary between the two regimes. These findings echo the empirical prevalence of division and (sexual) binary  reproduction in biological populations~\cite{Stearns}, while higher-order cooperative mechanisms~\cite{Miller2001,Levitan1991,CluttonBrock2002} are rare. 

\section*{Acknowledgments}
The research of O.V. is supported by grants from the NSF (DMS-2235451) and Simons Foundation (MPS-NITMB-00005320) to the NSF-Simons National Institute for Theory and Mathematics in Biology (NITMB).
The research of B.M. is supported by the Israel Science Foundation (Grant No.\ 1579/25).

\appendix

\section{Hydrodynamic solver}
\label{app:pde}

Assuming for simplicity  the  $x\to -x$ symmetry, we solve the HD free-boundary problem~\eqref{eq:pde}--\eqref{eq:mass} on the rescaled half-domain $y = x/L(t) \in [0,1]$. Defining $W(y,t) = L(t)\,u(L\,y,t)$, the transformed equation reads
\begin{equation}\label{eq:pde_rescaled}
  \dot{W} = \frac{1}{L^2}\,W_{yy}
    + \frac{\dot{L}}{L}\bigl(W + y\,W_y\bigr)
    + c\,\frac{W^k}{L^{k-1}}\,,
\end{equation}
with $W_y(0,t) = 0$ (symmetry), $W(1,t) = 0$ (absorbing boundary), and $\int_0^1 W\,\dd y = 1/2$ (mass constraint). For $k=3$ the coefficient is $c = \alpha$; for $k\ne 3$ the equation is rendered parameter-free by rescaling the spatial coordinate by
$\ell$.

The spatial domain is discretized on a uniform grid $y_j = j\,\Delta y$, $j=0,\ldots,N_y$, with central differences for diffusion and advection. At $y=0$ a ghost node enforces the Neumann condition; at $y=1$ the absorbing boundary sets $W=0$. Because the mass constraint determines~$L(t)$ implicitly, $L$ is treated  at each time step as an additional unknown. Time integration uses an implicit scheme (variable-order BDF with adaptive step size); at each step the resulting nonlinear algebraic system is solved by Newton iterations. The tridiagonal structure of the spatial operator is exploited to keep the cost per Newton step proportional to~$N_y$~\cite{SciPy2020}.

\section{Monte Carlo simulations}
\label{app:mc}

The lattice model is simulated with the Gillespie direct method~\cite{Gillespie1977}. Each step proceeds as follows.
\begin{enumerate}
\item Compute the total event rate
$W_{\mathrm{tot}} = W_{\mathrm{hop}} + W_{\mathrm{rxn}}$,
where $W_{\mathrm{hop}} = 2\mathcal{D}\,N$ is the total hopping rate and
$W_{\mathrm{rxn}} = \sum_j w_j$ with the per-site reproduction propensity
\begin{equation}\label{eq:propensity}
  w_j = \Lambda\,\frac{(n_j)_k}{k!}\,,
\end{equation}
where $(n)_k = n(n{-}1)\cdots(n{-}k{+}1)$ is the falling factorial and $n_j$ is the occupancy of site~$j$.

\item Draw the waiting time
$\delta t = -\ln U_1 / W_{\mathrm{tot}}$,
where $U_1$ is a uniform random number on $(0,1)$, and advance the clock.

\item With probability $W_{\mathrm{hop}}/W_{\mathrm{tot}}$, execute a
\emph{hop}: select a particle uniformly at random and move it to a
randomly chosen nearest neighbor.

\item Otherwise, execute a \emph{reproduction and removal}: select the
reproduction site~$j$ with probability $w_j/W_{\mathrm{rxn}}$, create a new
particle at~$j$, and simultaneously remove one particle from the
occupied site farthest from the origin. When $|j_L|$ and $|j_R|$ are compared, the site with the larger absolute distance is chosen; ties are broken arbitrarily. This coupled step conserves the total particle number~$N$ exactly.
\end{enumerate}

Initial conditions are deterministic discretizations of the analytical steady-state density profile (when available) or gamma-distributed density profiles across lattice sites. The simulator is implemented in C++, called from Python via a compiled extension module, with constant-time particle selection for hops and logarithmic-time reproduction-site sampling via a Fenwick tree.

\section{Analytical proof of instability for \texorpdfstring{$k>3$}{k>3}}
\label{app:instability}

Here we prove analytically that for $k > 3$ the steady state derived in Sec.~\ref{sec:steady} is linearly unstable. The notation and setup follow Sec.~\ref{sec:stability}. As shown there, all odd eigenvalues are strictly negative by a Sturm argument (the first positive zero of the odd unconstrained solution at \(\sigma=0\) is outside the interval \((0,L]\)). It remains to analyze the even sector, where the mass constraint~\eqref{eq:mass_pert} is the quantization condition.

Let $A_* = u_s(0)$ denote the peak density of the unit-mass steady state and $L_* = L(A_*)$ its half-width. For each real $\sigma$, define $v_\sigma$ as the solution of the initial-value problem
\begin{equation}\label{eq:v_sigma_ivp}
  D\,v_\sigma'' + \bigl(k\lambda\,u_s^{k-1}(x) - \sigma\bigr)\,v_\sigma = 0\,,
  \quad v_\sigma(0) = 1,\quad v_\sigma'(0) = 0\,,
  \quad 0 < x < L_*\,.
\end{equation}
The absorbing boundary condition at the perturbed edge, $v_\sigma(L_*) + \delta L\,u_s'(L_*) = 0$, merely determines the boundary displacement~$\delta L$ and does not impose an additional condition on $v_\sigma(L_*)$. The only remaining condition is mass conservation~\eqref{eq:mass_pert}, so the even constrained eigenvalues are precisely the zeros of
\begin{equation}\label{eq:shooting_J}
  J(\sigma) = \int_0^{L_*} v_\sigma(x)\,\dd x\,.
\end{equation}

To evaluate $J(0)$, consider the one-parameter family of compactly supported steady states $U_A(x)$ solving the ODE~\eqref{eq:ss_ode} for arbitrary peak density $U_A(0) = A$, without imposing the unit-mass condition. Let $L(A)$ be its half-width and $M(A) = 2\int_0^{L(A)} U_A(x)\,\dd x$ its total mass. Differentiating the steady-state equation $D\,U_A'' + \lambda\,U_A^k = 0$ with respect to $A$, we find that $W_A(x) := \partial_A U_A(x)$ satisfies the homogeneous linearized equation with $\sigma=0$:
\begin{equation}
  D\,W_A'' + k\lambda\,U_A^{k-1}\,W_A = 0\,.
\end{equation}
At $A = A_*$, the initial conditions are $W_{A_*}(0) = \partial_A [U_A(0)]|_{A_*} = 1$ and $W_{A_*}'(0) = \partial_A [U_A'(0)]|_{A_*} = 0$ (since $U_A'(0)=0$ for all $A$ by symmetry). These coincide with the initial conditions for $v_0$ in~\eqref{eq:v_sigma_ivp}, so $W_{A_*} = v_0$ by uniqueness. Applying Leibniz's rule to $\frac{1}{2}M'(A_*) = \frac{d}{dA}\int_0^{L(A)} U_A\,\dd x\big|_{A_*}$, and noting that the boundary term at $x=L(A)$ vanishes because $U_A(L(A)) = 0$ for all $A$, we obtain
\begin{equation}\label{eq:J0_mass_slope}
  J(0) = \int_0^{L_*} W_{A_*}(x)\,\dd x
       = \tfrac{1}{2}\,M'(A_*)\,.
\end{equation}
Using the mass formula~\eqref{eq:mass_general}:
\begin{equation}
    M(A) = 2\sqrt{D(k+1)/(2\lambda)}\;I_2(k)\;A^{(3-k)/2}\,,
\end{equation}
we find $M'(A) = \tfrac{3-k}{2A}\,M(A)$, so that $\operatorname{sgn} J(0) = \operatorname{sgn} M'(A_*) = \operatorname{sgn}(3-k)$.

For $k > 3$ we therefore have $J(0) < 0$. On the other hand, $J(\sigma) > 0$ for all sufficiently large~$\sigma$: when $\sigma$ exceeds
$k\lambda\,u_s^{k-1}(0)$, Eq.~\eqref{eq:v_sigma_ivp} has the form $D\,v'' - c(x)\,v = 0$ with $c(x) > 0$. The solution of this equation with the initial conditions $v(0)=1$, $v'(0)=0$ is positive and convex on $(0,L_*)$, giving $J(\sigma) > 0$. Therefore, by the intermediate value theorem, there exists $\sigma_0^{(\mathrm{c})} > 0$ with $J(\sigma_0^{(\mathrm{c})}) = 0$. This positive even eigenvalue implies linear instability of the steady states for all $k > 3$.

\section{Logarithmic correction to the boundary position for \texorpdfstring{$k=4$}{k=4}}
\label{app:log_correction}

Here we derive the long-time asymptotic behavior $L(t)\sim\sqrt{2Dt\ln t}$ for the diffusive spreading regime of the $k=4$ free-boundary problem. Differentiating the mass constraint $\int_{-L}^L u\,\dd x = 1$ with respect to time and using the absorbing boundary condition $u(\pm L,t)=0$, we obtain
\begin{equation}\label{eq:app_mass_balance}
  -D\bigl[u_x(L,t) - u_x(-L,t)\bigr] = \lambda\int_{-L}^{L} u^4\,\dd x\,,
\end{equation}
that is the total absorbing flux through the boundaries exactly balances the nonlinear source.

At long times the bulk density approaches the unit-mass Gaussian,
\begin{equation}\label{eq:app_gaussian}
  u(x,t) \simeq \frac{1}{\sqrt{4\pi Dt}}\,\exp\!\left(-\frac{x^2}{4Dt}\right),
\end{equation}
since the term $\lambda u^4 = O(t^{-2})$ is locally subleading compared to the two other terms of Eq.~(\ref{eq:pde}), which both scale as  $O(t^{-3/2})$. Nevertheless, the source is globally important: its spatial integral determines the exponentially small boundary flux in Eq.~(\ref{eq:app_mass_balance}). Substituting~\eqref{eq:app_gaussian} into the right-hand side of~\eqref{eq:app_mass_balance}, we obtain
\begin{equation}\label{eq:app_source_integral}
  \lambda\int_{-\infty}^{\infty} u^4\,\dd x
  \simeq \frac{\lambda}{16\pi^{3/2}(Dt)^{3/2}}\,.
\end{equation}

Now we set $L(t) = a(t)\sqrt{Dt}$ with $a(t)$ slowly increasing, so that the boundaries at $x=\pm L(t)$ expand faster than the bulk of the population. The total absorbing flux is
\begin{equation}\label{eq:app_flux}
  -D\bigl[u_x(L,t) - u_x(-L,t)\bigr]
  \simeq \frac{a(t)}{2\sqrt{\pi}\,t}\,\exp\!\left(-\frac{a^2(t)}{4}\right).
\end{equation}
Equating~\eqref{eq:app_flux} and~\eqref{eq:app_source_integral}, we obtain
\begin{equation}\label{eq:app_balance}
  a(t)\,\exp\!\left(-\frac{a^2(t)}{4}\right)
  \simeq \frac{\lambda}{8\pi D^{3/2}\sqrt{t}}\,.
\end{equation}
Taking logarithms of both sides and retaining only the leading contribution on the left-hand side gives $a^2(t)/4 \simeq \tfrac{1}{2}\ln t$. Therefore,
\begin{equation}\label{eq:app_result}
  L(t) = a(t)\sqrt{Dt} \simeq \sqrt{2Dt\ln t}\,,
  \qquad t\to\infty\,.
\end{equation}

\bibliographystyle{elsarticle-num}
\bibliography{references}

\end{document}